\documentclass{ws-procs9x6}

\def\draftnote{\hfill\hbox to \trimwidth{\footnotesize 
Talk presented by Takeuchi at SCGT'02, December 13, 2002, Nagoya, Japan\hfill VPI-IPPAP-03-04\qquad}}%

\begin{document}

\title{The $W$ mass and the $U$ parameter}

\author{Tatsu~Takeuchi${}^{(1)}$\footnote{Presenting Author},
Will Loinaz${}^{(2)}$, Naotoshi Okamura${}^{(1)}$, and
L.~C.~R.~Wijewardhana${}^{(3)}$}
\address{${}^{(1)}$Institute for Particle Physics and Astrophysics\\
Physics Department, Virginia Tech, Blacksburg, VA 24061\\
${}^{(2)}$Department of Physics, Amherst College, Amherst MA 01002\\
${}^{(3)}$Department of Physics, University of Cincinnati, Cincinnati OH 45221-0011}

\maketitle

\abstracts{
The $Z$-pole data from $e^+e^-$ colliders \cite{LEP/SLD:2002} 
and results from the NuTeV \cite{Zeller:2001hh} experiment at 
Fermilab can be brought into agreement if 
(1) the neutrino-$Z$ couplings were suppressed relative to the Standard Model (SM), and
(2) the Higgs boson were much heavier than suggested by SM global fits \cite{LEP/SLD:2002}.
However, increasing the Higgs boson mass will move the
theoretical value of the $W$ mass away from its experimental value.
A large and positive $U$ parameter becomes necessary to account for 
the difference.  We discuss what type of new physics may lead to 
such values of $U$.
}


\section{The NuTeV Anomaly and Neutrino Mixing} 

The NuTeV \cite{Zeller:2001hh} experiment at Fermilab 
has measured the ratios of neutral to charged current events in 
muon (anti)neutrino--nucleon scattering:
\begin{eqnarray}
R_\nu 
& = & \dfrac{ \sigma(\nu_\mu N \rightarrow \nu_\mu X) }
            { \sigma(\nu_\mu N \rightarrow \mu^-   X) }
\;=\; g_L^2 + r g_R^2\;, \cr
R_{\bar{\nu}}
& = & \dfrac{ \sigma(\bar{\nu}_\mu N \rightarrow \bar{\nu}_\mu X) }
            { \sigma(\bar{\nu}_\mu N \rightarrow \mu^+         X) }
\;=\; g_L^2 + \dfrac{g_R^2}{r}\;,
\end{eqnarray}
where
\begin{equation}
r = \dfrac{ \sigma( \bar{\nu}_\mu N \rightarrow \mu^+ X) }
          { \sigma( \nu_\mu       N \rightarrow \mu^- X) }
\sim \frac{1}{2}\;,
\end{equation}
and has determined the parameters $g_L^2$ and $g_R^2$ \cite{LlewellynSmith:ie}
to be
\begin{eqnarray}
g_L^2 & = & 0.30005 \pm 0.00137\;, \cr
g_R^2 & = & 0.03076 \pm 0.00110\;.
\label{nutev}
\end{eqnarray}
The Standard Model (SM) predictions of these parameters 
based on a global fit to non-NuTeV data, cited as 
$[g_L^2]_\mathrm{SM}=0.3042$ and 
$[g_R^2]_\mathrm{SM}=0.0301$ in Ref.~[\refcite{Zeller:2001hh}],
differ from the NuTeV result by $3\sigma$ in $g_L^2$.
This disagreement between NuTeV and the 
SM (as determined by non-NuTeV data) is sometimes referred to
as the NuTeV ``anomaly'' \cite{Davidson:2001ji}.

Note that the NuTeV value of $g_L^2$ is \textit{smaller} than the SM 
prediction.  In terms of the ratios $R_\nu$ and $R_{\bar{\nu}}$,  
this means that the neutral current events were not
as numerous as predicted by the SM when compared to the charged
current events.  
In addition, the $Z$ invisible width measured at $e^+e^-$ colliders, 
\begin{equation}
\Gamma_\mathrm{inv}/\Gamma_\mathrm{lept}
= 5.942 \pm 0.016 \;,
\end{equation}
is $2\sigma$ below the SM prediction of
$[\Gamma_\mathrm{inv}/\Gamma_\mathrm{lept}]_\mathrm{SM} = 5.9736 \pm 0.0036$
\cite{LEP/SLD:2002}.  
Both these observations seem to suggest that the neutrino
couplings to the $Z$ boson are suppressed.

Suppression of the $Z\nu\nu$ couplings occurs most naturally in
models which mix the neutrinos with heavy gauge singlet states \cite{mix}.
For instance, if the $SU(2)_L$ active neutrino $\nu_L$
is a linear combination of two mass eigenstates with
mixing angle $\theta$, 
\begin{equation}
\nu_L = (\cos\theta) \nu_\mathrm{light} + (\sin\theta) \nu_\mathrm{heavy}\;,
\end{equation}
then the $Z\nu\nu$ coupling will be suppressed by a factor of
$\cos^2\theta$ provided the heavy state is too massive to be created
in the interaction.  The $W\ell\nu$ coupling
will also be suppressed by a factor of $\cos\theta$.

In general, if the $Z\nu\nu$ coupling of a particular neutrino 
flavor is suppressed by a factor of $(1-\varepsilon)$, 
then the $W\ell\nu$ coupling of the same flavor will
be suppressed by a factor of $(1-\varepsilon/2)$.
For the sake of simplicity, 
assume that the suppression parameter $\varepsilon$ 
is common to all three generations. 
The theoretical values of $R_\nu$ and $R_{\bar{\nu}}$
are then reduced by a factor of $(1-\varepsilon)$, since their numerators
are suppressed over their denominators, while the invisible width of 
the $Z$ is reduced by a factor of $(1-2\varepsilon)$.
Thus, neutrino mixing could, in principle, provide an 
explanation for both the NuTeV and invisible width discrepancies.

At this point, we recall that one of the inputs used to calculate
SM predictions is the Fermi constant $G_F$, which is extracted
from the muon decay constant $G_\mu$.
The suppression of the $W\ell\nu$ couplings leads to the correction
\begin{equation}
G_F = G_\mu (1+\varepsilon)\;,
\label{GFshift}
\end{equation}
which would affect \textit{all} SM predictions.
One might worry that a shift in $G_F$ would destroy the excellent
agreement between the SM and the majority of the $Z$-pole data.
However, since the Fermi constant always appears multiplied by the
$\rho$-parameter in neutral current amplitudes, the shift can be compensated
by the introduction of the $T$ parameter \cite{LOTW1}, leaving 
$\rho G_F$ unaffected.
The suppression of $Z\nu\nu$ couplings together with
oblique corrections from new physics could thus reconcile the
NuTeV result with the $Z$-pole data.

\section{The Fits}

\begin{table}[t]
\ttbl{10cm}{The observables used in this analysis.
The SM predictions are for inputs of
$M_\mathrm{top}=174.3\,\mathrm{GeV}$,
$M_\mathrm{Higgs}=115\,\mathrm{GeV}$,
$\alpha_s(M_Z) = 0.119$, and
$\Delta\alpha_\mathrm{had}^{(5)} = 0.02761$.\vspace*{12pt}}
{\begin{tabular}{ccc}
\hline
Observable & SM prediction & Measured Value \\
\hline
$\Gamma_\mathrm{lept}$ & 
$83.998$~MeV & 
$83.984\pm 0.086$~MeV \\
$\Gamma_\mathrm{inv}/\Gamma_\mathrm{lept}$ &
$5.973$ &
$5.942 \pm 0.016$ \\
$\sin^2\theta_\mathrm{eff}^\mathrm{lept}$ & 
$0.23147$ &
$0.23148\pm 0.00017$ \\
$g_L^2$ & 
$0.3037$ &
$0.3002 \pm 0.0012$ \\
$g_R^2$ &
$0.0304$ &
$0.0310 \pm 0.0010$ \\
$M_W$ &
$80.375$ &
$80.449\pm 0.034$~GeV \\
\hline
\end{tabular}
\label{INPUTS}}
\end{table}

To test this idea, 
we fit the $Z$-pole, NuTeV,
and $W$ mass data with the oblique correction parameters 
$S$, $T$, $U$ \cite{Peskin:1990zt} and the $Z\nu\nu$ coupling suppression 
parameter $\varepsilon$.  
Table~\ref{INPUTS} comprises the six observables used in our
fit and their SM predictions.  
The details of the analysis are presented in Ref.~[\refcite{LOTW1}]; 
here, we merely outline our results.

That oblique corrections alone cannot account for the NuTeV
anomaly is established with a fit using only the oblique correction
parameters $S$, $T$, and $U$.
Taking $M_\mathrm{top} = 174.3\,\mathrm{GeV}$,
$M_\mathrm{Higgs} = 115\,\mathrm{GeV}$ as the reference SM, we obtain
\begin{eqnarray}
S & = & -0.09\pm 0.10\;, \cr
T & = & -0.13\pm 0.12\;, \cr
U & = & \phantom{-}0.32\pm 0.13\;.
\end{eqnarray}
The quality of the fit is unimpressive: $\chi^2=11.3$ for $6-3=3$ degrees of freedom. 
The preferred region on the $S$-$T$ plane is shown in Fig.~1a.
As is evident from the figure, there is no region where the $1\sigma$ bands
for $\Gamma_\mathrm{lept}$, $\sin^2\theta_\mathrm{eff}^\mathrm{lept}$,
and $g_L^2$ overlap.

\begin{figure}[t]
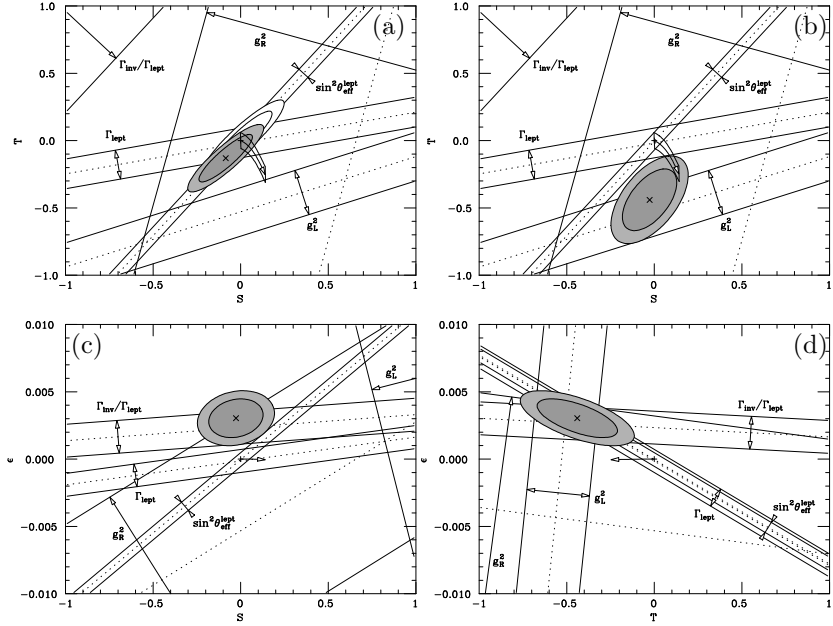

\begin{center}
\unitlength=1mm
\begin{picture}(110,35)(0,0)
\put(48,36){(a)}
\put(104,36){(b)}
\epsfxsize=5.5cm
\epsfbox[40 25 330 240]{fig1a.ps}
\epsfxsize=5.5cm
\epsfbox[40 25 330 240]{fig1b.ps}
\end{picture}
\begin{picture}(110,42)(0,0)
\put(8,36){(c)}
\put(104,36){(d)}
\epsfxsize=5.5cm
\epsfbox[40 25 330 240]{fig1c.ps}
\epsfxsize=5.5cm
\epsfbox[40 25 330 240]{fig1d.ps}
\end{picture}
\caption{The fit to the data with only $S$ and $T$ (a), and with
$S$, $T$, and $\varepsilon$ (b),(c),(d). 
The bands associated with each observable show the $1\sigma$ limits.
(The $M_W$ band is not shown.)
The shaded ellipses show the 68\% and 90\% confidence contours. 
The unshaded ellipses partially hidden behind the shaded ones 
in (a) show the contours when only the $Z$-pole data is used.  
The origin is the reference SM with $M_\mathrm{top}=174.3\,\mathrm{GeV}$ and
$M_\mathrm{Higgs}=115\,\mathrm{GeV}$.  The curved arrow attached to
the origin indicates the path along which the SM point will move when the
Higgs mass is increased from 115~GeV to 1~TeV.}
\end{center}
\end{figure}

Next, we fit using $S$, $T$, $U$, and $\varepsilon$.
The reference SM is $M_\mathrm{top} = 174.3\,\mathrm{GeV}$,
$M_\mathrm{Higgs} = 115\,\mathrm{GeV}$ as before.
The result is
\begin{eqnarray}
S & = & -0.03\pm 0.10\;, \cr
T & = & -0.44\pm 0.15\;, \cr
U & = & \phantom{-}0.62\pm 0.16\;, \cr
\varepsilon & = & 0.0030 \pm 0.0010\;.
\end{eqnarray}
The quality of the fit is improved dramatically to $\chi^2=1.17$ for
$6-4=2$ degrees of freedom.
The preferred regions in the $S$-$T$, $S$-$\varepsilon$, and $T$-$\varepsilon$
planes are shown in Figs.~1b through 1d.
As anticipated, inclusion of both oblique corrections and $\varepsilon$
results in an excellent fit to both the $Z$-pole and NuTeV data.

\section{Heavy Higgs and the $W$ Mass}

\begin{figure}[ht]
\begin{center}
\unitlength=1mm
\begin{picture}(110,35)(0,0)
\put(48,36){(a)}
\put(105,8){(b)}
\epsfxsize=5.7cm
\epsfbox[35 30 365 270]{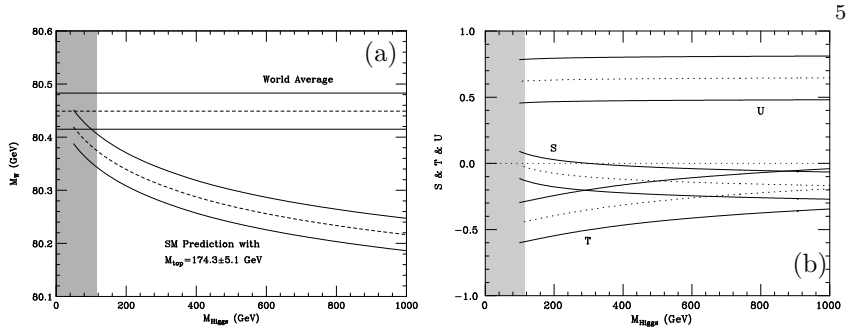}
\epsfxsize=5.5cm
\epsfbox[40 20 335 240]{fig2b.ps}
\end{picture}
\caption{The $M_\mathrm{Higgs}$ dependence of the SM prediction of 
$M_W$ (a), and the $1\sigma$ limits on $S$, $T$, and $U$ (b).
The Higgs mass in the shaded region is excluded by direct searches.}
\label{WMASSFIG}
\end{center}
\end{figure}

What type of new physics would provide the values of 
the oblique correction parameters and $\varepsilon$ preferred by the fit?
The value of $\varepsilon$ implies a largish mixing angle between the
light active and heavy sterile states.
In Ref.~[\refcite{LORTW2}], we discuss how such mixings can be realized 
within the seesaw framework.
For the oblique corrections,
the limits on $S$ permit it to have either sign, while $T$ is
constrained to be negative by $3\sigma$.  Few models of new physics 
are available which predict a negative $T$ \cite{Bertolini:1990ek}.

A heavy SM Higgs provides a simple starting point.
Recall that the effect of a SM Higgs heavier than our reference value 
(here chosen to be $115\,\mathrm{GeV}$) is manifested as shifts in the 
oblique correction parameters.
The approximate expressions for these shifts are
\begin{eqnarray}
S_\mathrm{Higgs} & \approx & \frac{1}{6\pi}
     \ln\left(M_\mathrm{Higgs}/M_\mathrm{Higgs}^\mathrm{ref}\right)\;, \cr
T_\mathrm{Higgs} & \approx & -\frac{3}{8\pi c^2}
     \ln\left(M_\mathrm{Higgs}/M_\mathrm{Higgs}^\mathrm{ref}\right)\;, \cr
U_\mathrm{Higgs} & \approx & 0\;.
\label{HiggsSTU}
\end{eqnarray}
Thus increasing the Higgs mass generates a negative $T$.  
Indeed, we have shown in Ref.~[\refcite{LOTW1}] 
that the $Z$-pole and NuTeV observables can be fit by 
$\varepsilon$ alone if the Higgs boson is as heavy
as a few hundred GeV.
Since the Higgs boson has not been found in the $\sim 80$~GeV range 
preferred by the SM global fit \cite{LEP/SLD:2002}, the
prospect that the data prefer a heavier Higgs is
actually welcome \cite{Peskin:2001rw,Chanowitz:2001bv}.

However, as shown in Fig.~2a, 
a heavier Higgs will lower the SM prediction of the $W$ mass,
shifting it away from the experimental
value.  We would like to point out that although the experimental value of
the $W$ mass differs from the SM global fit (with $M_\mathrm{Higgs}\sim 80$~GeV)
by only $1.7\sigma$ \cite{LEP/SLD:2002}, if the Higgs mass is raised to
its lower bound of 115~GeV, the difference is $2.2\sigma$.
As the experimental error on the $W$ mass decreases and the lower bound
on the Higgs mass increases, the $W$ mass may become
the next `anomaly' to be confronted.

Regardless the actual mass of the Higgs, our fits indicate that
the presence of $Z\nu\nu$ suppression demands a large and positive $U$ parameter 
to account for the $W$ mass. (See Fig.~2b.)
What new physics predicts a small $T$ and a large $U$?
One possibility is that the $U$ parameter is enhanced
by the formation of bound states at new particle thresholds.
Expressing $T$ and $U$ as dispersion integrals over
spectral functions gives
\begin{eqnarray}
T & \propto & \int_{s_\mathrm{thres}}^\infty \frac{ds}{s}
              \left[ \mathrm{Im}\Pi_\pm(s) - \mathrm{Im}\Pi_0(s)
              \right]\;, \cr
U & \propto & \int_{s_\mathrm{thres}}^\infty \frac{ds}{s^2}
              \left[ \mathrm{Im}\Pi_\pm(s) - \mathrm{Im}\Pi_0(s)
              \right]\;,
\end{eqnarray}
using the notation of Ref.~[\refcite{Takeuchi:1994ym}].
Because of the extra negative power of $s$ in its integrand,
$U$ is more sensitive to the threshold enhancement than $T$.
Indeed, it has been shown in Ref.~[\refcite{Takeuchi:1994ym}] that
threshold effects do not enhance the $T$ parameter.
This could be an indication that technicolor theories are the most
promising candidates.  Thus, technicolor theories which were
killed by the $S$ parameter \cite{Peskin:1990zt}
could be resurrected by the $U$ parameter.

\section*{Acknowledgments}

We thank Lay Nam Chang, Michael Chanowitz, 
Michio Hashimoto, Randy Johnson, Alex Kagan, Kevin McFarland, 
Mihoko Nojiri, Jogesh Pati, and Mike Shaevitz
for helpful discussions and communications.
This research was supported in part by the U.S. Department of Energy, 
grants DE--FG05--92ER40709, Task A (T.T. and N.O.),
and DE-FG02-84ER40153 (L.C.R.W.).



\end{document}